\documentclass[twocolumn]{revtex4}

\usepackage{graphicx}
\usepackage{dcolumn}
\usepackage{amsmath, amssymb}
\usepackage{color}

\begin{document}

\title{Unconventional and anomalous magnetic field distribution 
in a bilayer superconductor with geometric constraints
}

\author{Takashi Yanagisawa
}

\affiliation{Electronics and Photonics Research Institute, Advanced Engineering
Research Institute,
National Institute of Advanced Industrial Science and Technology (AIST),
1-1-1 Umezono, Tsukuba 305-8568, Japan\\
}


\begin{abstract}
We investigate the magnetic field distribution in multi-component superconductors.
We examine a layered superconductor and a two-component one-layer superconductor.
We evaluate the field distribution in the presence of a half-flux quantum
vortex with a kink structure in the phase space of gap functions.
We also examine the magnetic field distribution of a knot soliton which is
formulated in a two-component superconductor.
We investigate the effect of geometric constraints for multi-component
superconductors, where
the geometric constraint means that the system is compactified in one direction
so that the current in this direction becomes vanishingly small.  This corresponds
to the gauge fixing in this direction.
An unconventional magnetic field distribution takes place; here the unconventional
means that the  magnetic
field is screened incompletely which may be called the anomalous Meissner effect. 
We argue that this anomalous behavior creates a massless gauge field.
\end{abstract}


\maketitle

\section{Introduction}

Multi-component superconductivity provides significant phenomena
concerning magnetic properties in superconductors and has been studied
intensively.\cite{mos59,suh59,per62,kon63,sta10,tan10a,tan10b,dia11,yan12,hu12,sta12,
pla12,mai13,wil13,yer07,gur06,nic05,yer23,kut21,sam12,yer21,vak12}
Interesting phenomena in multi-component superconductors are the following:
time-reversal symmetry 
breaking,\cite{sta10,tan10a,tan10b,dia11,yan12,hu12,sta12,pla12,mai13,wil13,gan14,yer15}
the emergence of massless modes,\cite{yan13,lin12,kob13,koy14,yan14,tan15,sha02,yan17c}
the existence of fractional-flux quantum 
vortices,\cite{yan12,izy90,vol09,tan01,bab02,col05,blu06,gor07,chi10,kup11,gar11,pin12,gar13,
smi05,yan18c,tan18b,tan18,hay19,nag19,yan19,yan20} and also unconventional isotope
effect,\cite{cho09,shi09,yan09} and a new type of superconductor.\cite{mos09}
These significant properties are mainly originate from multi-phase structures of
gap functions.  For example, the existence of a fractional-flux quantum vortex is due
to a non-trivial kink structure of the phase difference between gaps. 

In this paper we examine the magnetic property of multi-component superconductors.
We consider a superconducting bilayer and also a superconducting layer with 
two-component order parameters.  The phase difference of the order parameters
forms a soliton-like (kink) structure.  
A half-flux quantum vortex (HFQV or HQV) exists with this kink. 
We discuss the magnetic field distribution in the presence of a HFQV
associated with a kink (soliton).  A structure in the phase space gives a
contribution to the magnetic field in superconductors.
The field equation for HFQV and a kink attached to it is solved to determine the
field distribution.

The derivative of the phase $\varphi$ contributes to the vector potential.
$\nabla\varphi$, however, gives nothing to the magnetic field ${\bf B}$ when
$\varphi$ has no singularity because of the identity ${\rm rot grad}\varphi=0$.
We investigate a multi-component superconductor where $\nabla\varphi$ gives non-zero
contributions to ${\bf B}$.  For this purpose, 
we examine the effect of geometrical constraints on two-component superconductors.
There is a possibility that non-trivial behavior of phase difference modes may
lead to an anomalous behavior of the Meissner effect that may be called the
incomplete or anomalous Meissner effect (IME or AME).
In a superconductor this indicates a phenomenon that the screening of
magnetic filed is incomplete and the magnetic field penetrates into a
superconductor partially.
There appears a massless mode with this anomalous behavior of magnetic field.
This shows a possibility that a massless mode exists even in a two-component
superconductor.
We show that we have an unusual magnetic field distribution due to geometric constraints
in a bilayer superconductor.
In a one-layer two-component superconducting disk, the constraint condition is
approximately satisfied automatically and an unusual magnetic field contribution is small.

The Higgs mechanism is an important phenomenon in a coupled scalar-gauge field
system.  
The Meissner effect in superconductors is the Higgs mechanism in a U(1) gauge theory.
This mechanism gives the mass to gauge fields as a result of
a spontaneous symmetry breaking.
The incomplete Meissner effect implies an incomplete Higgs mechanism.
The incomplete Higgs state means that some components of gauge field remain massless
when the gauge symmetry is spontaneously broken.

The paper is organized as follows.  In Section 2 we present a model free energy
and a model of a bilayer superconductor.
In Section 3 we consider the magnetic field distribution in a bilayer superconductor.
We show the field equation and calculate the field distribution.
In Section 4 we discuss the unconventional Meissner state and a gapless mode in a 
compactified superconducting bilayer.
In the last Section we give a summary.

\section{Multi-component gauged scalar fields}

\subsection{Ginzburg-Landau free energy}
 
We examine a multi-component scalar model coupled to the gauge field.
The model is that for superconductors with multi-component order parameters
in a magnetic field.
The free energy density is written in the form
\begin{align}
f &= \sum_j \alpha_j|\psi_j|^2 +\frac{1}{2}\sum_j\beta_j|\psi_j|^4
-\sum_{i<j}J_{ij}(\psi_i^*\psi_j+\psi_j^*\psi_i),
\nonumber\\
&~ +\sum_j\frac{\hbar^2}{4m_j}\Big|\left(\nabla-i\frac{e^*}{\hbar c}{\bf A}\right)
\psi_j\Big|^2 +\frac{1}{8\pi}(\nabla\times{\bf A})^2
\end{align}
where $\psi_j$ $(j=1,2,\cdots)$ are the order parameters and we set $e^*=2e$ for
the electron charge $e(<0)$.
 
The free energy $f$ is not invariant under the gauge transformation because of
the Josephson term $J_{ij}$.  The Josephson term is not invariant even for
a global gauge transformation. 
The characteristic feature of multi-component models in the presence of
the Josephson coupling is the existence of domain structure of phases of gap 
functions.
The domain structure means a soliton-like structure in the phase space
of gap functions.\\

\subsection{Multi-component model of a multi-layer superconductor}

A multi-layer superconductor is  a multi-component superconductor
when each layer is regarded as a one-component superconductor
(Fig.1 for a bilayer superconductor).
Each component of the superconductor exists in a different space and they
interact each other only through the Josephson coupling.
The magnetic vector potential ${\bf A}$ can be different in each space (layer)
since the order parameters $\psi_j$ ($j=1,2,\cdots$) can have different phases $\varphi_j$.

We consider two superconducting layers apart from the distance $s$ in the $x-y$ plane
and their thickness is $d$.
We assume that $d$ is small compared to the penetration depth $\Lambda$ which
is defined in Section 3.
We will consider the case where $s$ is also small compared to $\Lambda$:
\begin{equation}
d, s \ll \Lambda.
\end{equation}

\begin{figure}[htbp]
\begin{center}
  \includegraphics[height=2.7cm]{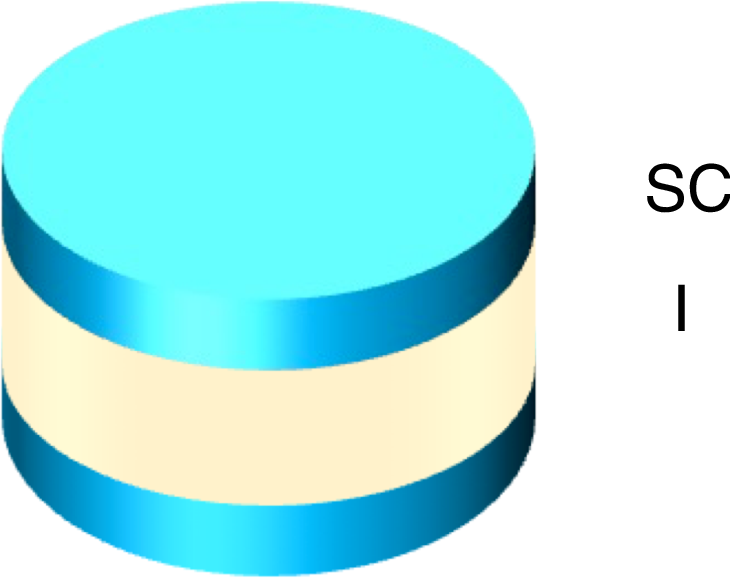}
\caption{A bilayer superconductor with an insulating layer between
two superconducting layers.
}
\label{fig1}       
\end{center}
\end{figure}

\section{Magnetic field in a bilayer superconductor}

\subsection{Field equations}

We have two regions $\Omega_1$ and $\Omega_2$ that denote layer 1 and layer 2,
respectively.  $\Omega_1$ is at $z=0$ and $\Omega_2$ is at $z=s$ where $s$ is
the distance between two layers.  The coordinate ${\bf x}$ is denoted as
${\bf x}=({\bf r},z)$ where ${\bf r}=(x,y)$ is the two-dimensional coordinate.
The vector potential ${\bf A}$ reads
\begin{eqnarray}
{\bf A}= \left\{
\begin{array}{c}
{\bf A}_1~~~{\rm in}~~\Omega_1 \\
{\bf A}_2~~~{\rm in}~~\Omega_2 . \\
\end{array}
\right. 
\end{eqnarray}
We define the phases $\varphi_j$ ($j=1,2,\cdots$) of gap functions as
\begin{equation}
\psi_j = e^{i\varphi_j}|\psi_j|= e^{i\varphi_j}\rho_j,
\end{equation}
where we set $\rho_j=|\psi_j|$.
The free energy is written as
\begin{align}
f &= \sum_j \alpha_j\rho_j^2 +\frac{1}{2}\sum_j\beta_j\rho_j^4
+ \sum_j\frac{\hbar^2}{4m_j}\left(\nabla\rho_j\right)^2 \nonumber\\
&~ +\sum_j\frac{\hbar^2}{4m_j}\rho_j^2\left(\nabla\varphi_j\right)^2 
-\frac{2e^*}{\hbar c}\sum_j\frac{\hbar^2}{4m_j}\rho_j^2\nabla\varphi_j\cdot{\bf A}_j
\nonumber\\
&~ -2J_{12}\rho_1\rho_2\cos\varphi \nonumber\\
&~ +\left(\frac{e^*}{\hbar c}\right)^2\frac{\hbar^2}{4}\left(\frac{\rho_1^2}{m_1}{\bf A}_1^2
+\frac{\rho_2^2}{m_2}{\bf A}_2^2 \right)
+\frac{1}{8\pi}\left(\nabla\times{\bf A}\right)^2,
\end{align}
where the phase difference $\varphi$ is defined as
\begin{equation}
\varphi({\bf r})= \varphi_1({\bf r},z=0)-\varphi_2({\bf r},z=s).
\end{equation}

The field equation for ${\bf A}$ in each layer reads
\begin{equation}
-\nabla\times (\nabla\times {\bf A}_j)= \frac{1}{\lambda_j^2}{\bf A}_j
+\frac{1}{\lambda_j^2}\frac{\phi_0}{2\pi}\nabla\varphi_j,
\end{equation}
for $j=1,2$ where $\lambda_j$ is the penetration depth in $j$-th layer given as
\begin{equation}
\lambda_j^{-2}= \frac{4\pi(e^*)^2}{2m_jc^2}\rho_j^2,
\end{equation}
and $\phi_0$ is the unit quantum flux: $\phi_0=hc/|e^*|$.  
The equations for $\varphi_j$ are given by
\begin{eqnarray}
&&\frac{\hbar^2\rho_j^2}{2m_j}
\left(\nabla^2\varphi_j-\frac{e^*}{\hbar c}\nabla{\bf A}_j\right)
+\frac{\hbar^2}{2m_j}\left(\nabla\varphi_j-\frac{e^*}{\hbar c}{\bf A}_j\right)
\nabla\rho_j^2 \nonumber\\
&&~~~ +J_{12}\rho_1\rho_2\sin\varphi =0.
\end{eqnarray}
In this paper we neglect the spatial dependence of $\rho_j$ since we are
interested in the physics originating from phase structure.

The equations for ${\bf A}_j$ show that ${\bf A}$ in each layer has the mass
$m_j=\lambda_j^{-1}$, which means that the Meissner effect occurs in each layer.
We discuss later that the Meissner effect will be modified to be incomplete 
when there is a geometric constraint for superconducting layers.

\subsection{Magnetic field in the presence of a half-flux quantum vortex with a kink} 

\subsubsection{HFQV in two equivalent layers}

We investigate the distribution of the magnetic field in the presence of a
half-flux quantum vortex (HFQV) in a superconducting bilayer.
This is an extension of that in a superconducting layer\cite{pea64} to the case
of a superconducting bilayer.
We assume that two superconductors are identical, that is, we have $\rho_1=\rho_2$
and $m_1=m_2$ in this subsection.
The width of the film is $d$ and $s$ denotes the distance between two layers.
In our model the HFQV exists accompanied with a kink (soliton) structure in 
the phase space of gap functions.
We write the phases $\varphi_j$ as
\begin{equation}
\varphi_1({\bf r},z=0)= \Phi+\frac{1}{2}\varphi,~~ \varphi_2({\bf r},z=s)= \Phi-\frac{1}{2}\varphi,
\end{equation}
where ${\bf r}\equiv (x,y)$ indicates the two-dimensional coordinate.  The total phase $\Phi$ is
defined by
\begin{equation}
\Phi({\bf r})= \frac{1}{2}\left(\varphi_1({\bf r},z=0)+\varphi_2({\bf r},z=s)\right).
\end{equation}

We use the Coulomb gauge $\nabla\cdot{\bf A}=0$, then
the vector potential ${\bf A}$ satisfies the 
following equation:
\begin{align}
\nabla^2{\bf A}({\bf r},z) &= \frac{2}{\Lambda}\delta(z)\left( {\bf A}({\bf r},z)
+\frac{\phi_0}{2\pi}\nabla\varphi_1 \right) \nonumber\\
&~ +\frac{2}{\Lambda}\delta(z-s)\left( {\bf A}({\bf r},z)
+\frac{\phi_0}{2\pi}\nabla\varphi_2\right),
\end{align}
where we put
\begin{equation}
\frac{2}{\Lambda}= \frac{d}{\lambda_1^2}=\frac{d}{\lambda_2^2}.
\end{equation}

The total phase $\Phi$ represents an HFQV that is given by
$\Phi= \frac{1}{2}\phi$, where $\phi= \tan^{-1}(y/x)$.
We have
\begin{equation}
\frac{\phi_0}{2\pi}\nabla\Phi= \frac{\phi_0}{4\pi}
\left( \frac{y}{x^2+y^2},-\frac{x}{x^2+y^2},0 \right).
\end{equation}
We adopt that the kink exists on the positive axis $x>0$ and the phase difference
$\varphi$ changes from 0 to $2\pi$ when crossing the kink from negative ($y<0$) to 
positive ($y>0$) regions.  
We introduce the Fourier transformation
\begin{equation}
{\bf A}({\bf r},z)= \int\frac{d^2q}{(2\pi)^2}\frac{dq_z}{2\pi}{\bf A}({\bf q},q_z)
e^{i{\bf q}\cdot{\bf r}+iq_zz},
\end{equation}
where we set ${\bf q}=(q_x,q_y)$.  
The Fourier transform of $\nabla\varphi({\bf r})$ is defined as
\begin{equation}
{\bf F}_{\Phi}({\bf q})\equiv \int dxdy\nabla\phi({\bf r})e^{-i{\bf q}\cdot{\bf r}} 
= 2\pi i\left( \frac{q_y}{q^2},-\frac{q_x}{q^2},0 \right),
\end{equation}
where $q=|{\bf q}|$.
We define the Fourier transform of $\nabla\varphi$ as
\begin{equation}
{\bf F}_{\varphi}({\bf q})\equiv \int dxdy\nabla\varphi({\bf r})e^{-i{\bf q}\cdot{\bf r}}. 
\end{equation}
By using the delta function representation
$\delta(z)= (1/2\pi)\int_{-\infty}^{\infty}dq_z e^{iq_zz}$,
we have
\begin{align}
{\bf A}({\bf q},q_z)&= -\frac{2}{\Lambda}\frac{1}{q^2+q_z^2}
\left( {\bf A}({\bf q},z=0)+\frac{\phi_0}{4\pi}{\bf F}_{\Phi}({\bf q})
+\frac{\phi_0}{4\pi}{\bf F}_{\varphi}({\bf q}) \right) \nonumber\\
&~ -\frac{2}{\Lambda}\frac{e^{-iq_zs}}{q^2+q_z^2} \left( {\bf A}({\bf q},z=s)
+\frac{\phi_0}{4\pi}{\bf F}_{\Phi}({\bf q})
-\frac{\phi_0}{4\pi}{\bf F}_{\varphi}({\bf q})\right).
\end{align}
This leads to
\begin{align}
{\bf A}({\bf q},z) &= -\frac{1}{\Lambda q}e^{-q|z|}
\left( {\bf A}({\bf q},z=0)+\frac{\phi_0}{4\pi}{\bf F}_{\Phi}({\bf q})
+\frac{\phi_0}{4\pi}{\bf F}_{\varphi}({\bf q}) \right) \nonumber\\
&~ -\frac{1}{\Lambda q}e^{-q|z-s|}\left( {\bf A}({\bf q},z=s)
+\frac{\phi_0}{4\pi}{\bf F}_{\Phi}({\bf q})
-\frac{\phi_0}{4\pi}{\bf F}_{\varphi}({\bf q})\right).
\end{align}
From this equation, ${\bf A}({\bf r},z)$ at $z=0$ and $z=s$ are determined.
We obtain ${\bf A}({\bf r},z)$ for $z=0$ and $z=s$, after the Fourier 
transformation, as
\begin{align}
{\bf A}({\bf r},0)&= -\frac{\phi_0}{4\pi} \int\frac{d^2q}{(2\pi)^2}e^{i{\bf q}\cdot{\bf r}}
\bigg[ \frac{1+e^{-qs}}{\Lambda q+1+e^{-qs}}
{\bf F}_{\Phi}({\bf q})\nonumber\\
&~~~~~ +\frac{1-e^{-qs}}{\Lambda q+1-e^{-qs}}
{\bf F}_{\varphi}({\bf q}) \bigg],\\ 
{\bf A}({\bf r},s)&= -\frac{\phi_0}{4\pi} \int\frac{d^2q}{(2\pi)^2}e^{i{\bf q}\cdot{\bf r}}
\bigg[ \frac{1+e^{-qs}}{\Lambda q+1+e^{-qs}} {\bf F}_{\Phi}({\bf q}) \nonumber\\
& ~~~~~ -\frac{1-e^{-qs}}{\Lambda q+1-e^{-qs}} {\bf F}_{\varphi}({\bf q}) \bigg].
\end{align}
This results in
\begin{align}
{\bf A}({\bf r},z)&= -\frac{\phi_0}{4\pi} \int\frac{d^2q}{(2\pi)^2}e^{i{\bf q}\cdot{\bf r}}
\bigg[ \frac{e^{-q|z|}+e^{-q|z-s|}}{\Lambda q+1+e^{-qs}}{\bf F}_{\Phi}({\bf q}) \nonumber\\
& ~~~~~~~~ + \frac{e^{-q|z|}-e^{-q|z-s|}}{\Lambda q+1-e^{-qs}} {\bf F}_{\varphi}({\bf q})
\bigg].
\end{align}

\subsubsection{Field distribution in the presence of HFQV}

We divide ${\bf A}$ into radial and angular components as
\begin{equation}
{\bf A}= A_r{\bf e}_r+A_{\phi}{\bf e}_{\phi},
\end{equation}
where ${\bf e}_r=(\cos\phi,\sin\phi)$ and ${\bf e}_{\phi}=(-\sin\phi,\cos\phi)$
are unit vectors in the polar coordinate for ${\bf r}=r(\cos\phi,\sin\phi)$.
Let $\theta$ be the angle between vectors ${\bf r}$ and ${\bf q}$, so that
we write ${\bf q}= q(\cos(\theta+\varphi),\sin(\theta+\varphi))$.

The field ${\bf F}_{\varphi}({\bf q})$ indicates that generated by the kink
distribution $\varphi$.  Here we examine two cases where  
${\bf F}_{\varphi}({\bf q})= {\bf F}_{\Phi}({\bf q})$ and 
${\bf F}_{\varphi}({\bf q})\neq {\bf F}_{\Phi}({\bf q})$.
This depends on the spatial dependence of $\varphi$; we show two cases in Fig. 2.

\begin{figure}[htbp]
\begin{center}
  \includegraphics[height=6cm, angle=270]{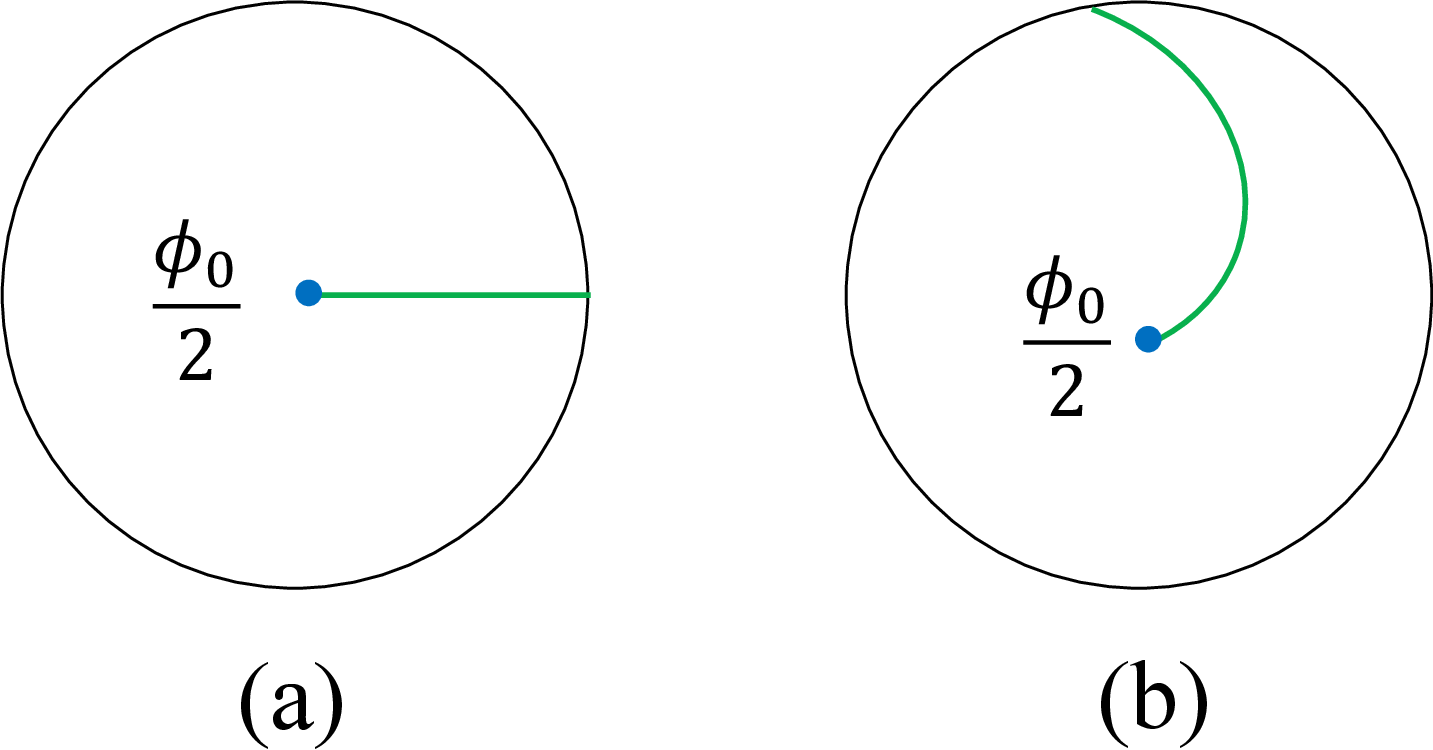}
\caption{Distributions of phase kinks: (a) there is a cut on the positive 
$x$-axis as shown in eq.(\ref{cut}) and (b) a cut has a spatial dependence as 
$y=f(x)$ for some real function $f(x)$.  In the case (b) we have
${\bf F}_{\varphi}({\bf q})\neq {\bf F}_{\Phi}({\bf q})$ in general. 
There is an HFQV at the center.
}
\label{fig2}       
\end{center}
\end{figure}

\paragraph{{\rm (A)} The case ${\bf F}_{\varphi}({\bf q})= {\bf F}_{\Phi}({\bf q})$}

In this case the vector potential agrees with that obtained for the model
where the unit flux exists in one layer and there is no flux in the other layer.
We use the representation
\begin{equation}
\varphi= -{\rm Im}\log(x+iy)+2\pi.
\label{cut}
\end{equation}
This results in
\begin{equation}
\frac{\phi_0}{4\pi}\nabla\varphi= \frac{\phi_0}{4\pi}
\left( \frac{y}{x^2+y^2},-\frac{x}{x^2+y^2},0 \right),
\end{equation}
and then we have ${\bf F}_{\varphi}({\bf q})= {\bf F}_{\Phi}({\bf q})$.

Since $A_{\phi}={\bf A}\cdot{\bf e}_{\phi}$, we have
\begin{align}
A_{\phi}({\bf r},z)&= \frac{\phi_0}{2\pi}\int_0^{\infty} dq
\bigg[ -e^{-q|z|}(\Lambda q+1)+e^{-q|z-s|-qs} \bigg] \nonumber\\
&~~~~~   \cdot\frac{1}{(\Lambda q+1)^2-e^{-2qs}}J_1(qr),
\end{align}
where $J_1(x)$ is the Bessel function of the first kind.
We have also $A_r({\bf r},z)=0$.
In the limit of small $s$, $A_{\phi}({\bf r},z)$ reduces to
\begin{equation}
A_{\phi}({\bf r},z)\simeq -\frac{\phi_0}{2\pi}\int_0^{\infty}dqe^{-q|z|}
\frac{1}{\Lambda q+2}J_1(qr).
\end{equation}
This formula is very similar to that for Pearl's vortex. 
We should note that
we have the constant 2 in the denominator $\Lambda q+2$ in the integrand.
This constant 2 is important because it represents that the vortex has
half-quantum flux $\phi_0/2$.  In fact, the total flux $\Phi_{total}$
in the disc of radius $r_{disc}$ at $z=0$ is given by
\begin{align}
\Phi_{total}(0)&= 2\pi \int_0^{r_{disc}}drrB_z({\bf r},0)
= -2\pi r_{disc} A_{\phi}(r_{disc},0)\nonumber\\
&\longrightarrow \frac{1}{2}\phi_0~~{\rm for}~~r_{disc}\gg \Lambda,
\end{align}
where the magnetic field $B_z$ is given as
\begin{equation}
B_z({\bf r},z)= \frac{1}{r}\frac{\partial A_r}{\partial\phi}
-\frac{1}{r}\frac{\partial}{\partial r}\left( rA_{\phi}(r,z)\right).
\end{equation}
Thus we have a half-flux quantum.

We show $A_{\phi}({\bf r},z)$ in units of $-\phi_0/(2\pi\Lambda)$ as a
function of $r/\Lambda$ in Fig. 3 where $z=0.1$ and $s=0.1, 0.5$ and 1.0.
Here the length is measured in units of $\Lambda$.
The Fig. 4 shows $A_{\phi}({\bf r},z)$ for $z=1.0$ and $s=0.1$, 0.5 and 1.0.
$-A_{\phi}$ has a peak for small $r/\Lambda$ when $z/\Lambda$ is small and decreases 
as $z$ increases.
We present the magnetic field $B_z({\bf r},z)$ in unit of $\phi_0/(2\pi\Lambda^2)$ 
in Fig. 5.

\begin{figure}[htbp]
\begin{center}
  \includegraphics[height=5.0cm]{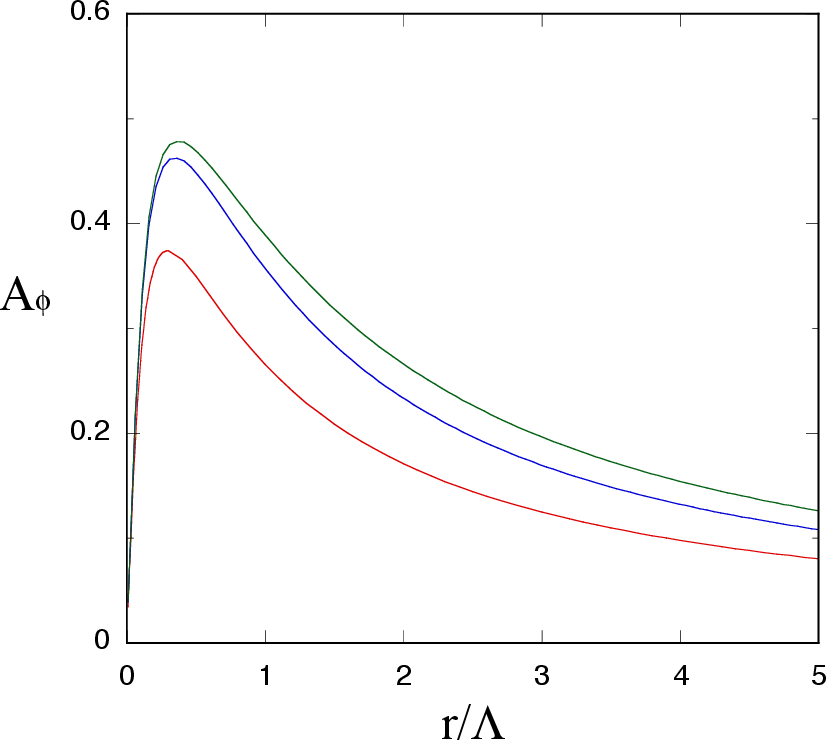}
\caption{$A_{\phi}$ in units of $-\phi_0/(2\pi\Lambda)$ as a function of 
$r/\Lambda$ for $z=0.1$ and $s=0.1$, 0.5
and 1.0 form the bottom in units of $\Lambda$.
}
\label{fig3}       
\end{center}
\end{figure}

\begin{figure}[htbp]
\begin{center}
  \includegraphics[height=5.0cm]{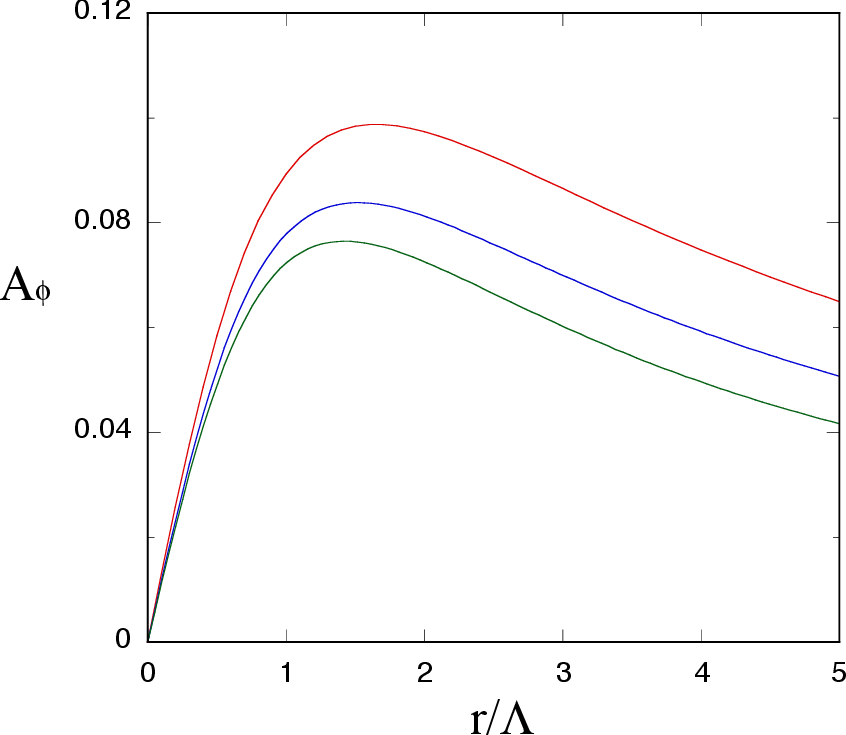}
\caption{$A_{\phi}$ in units of $-\phi_0/(2\pi\Lambda)$ as a function of 
$r/\Lambda$ for $z=1.0$ and $s=0.1$, 0.5
and 1.0 form the top in units of $\Lambda$.
}
\label{fig4}       
\end{center}
\end{figure}

\begin{figure}[htbp]
\begin{center}
  \includegraphics[height=5.0cm]{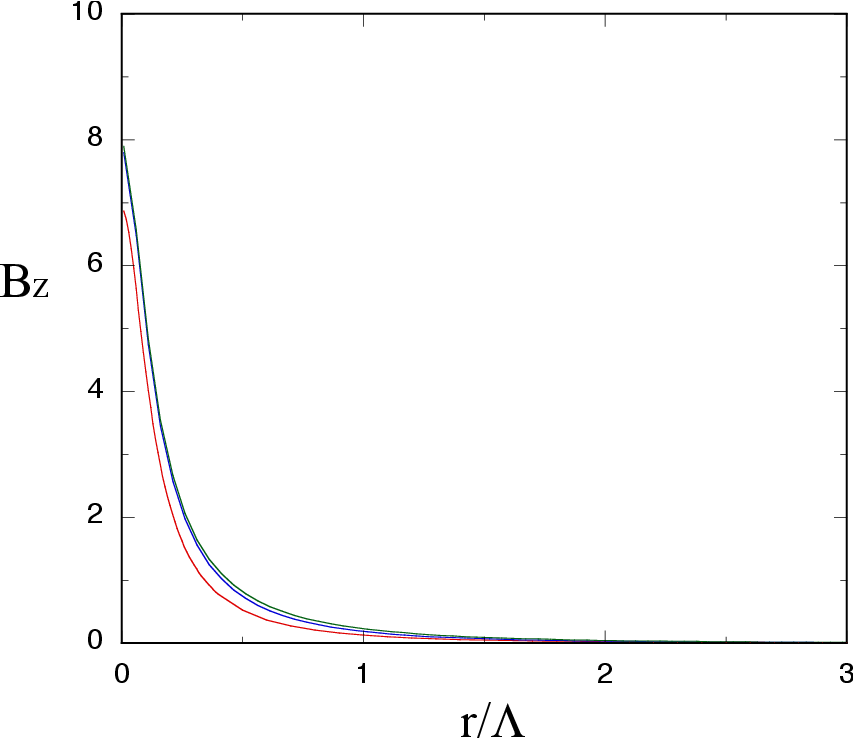}
\caption{$B_{z}$ in units of $\phi_0/(2\pi\Lambda^2)$ as a function of 
$r/\Lambda$ for $z=0.1$ and $s=0.1$, 0.5
and 1.0 form the bottom in units of $\Lambda$.
}
\label{fig5}       
\end{center}
\end{figure}

\paragraph{{\rm (B)} The case ${\bf F}_{\varphi}({\bf q})\neq {\bf F}_{\Phi}({\bf q})$}

In this case the vector potential does not necessarily agrees with that for
the one unit-flux model in contrast to the case A.
Here we take an alternative form of the phase difference $\varphi$.
Here we assume that $s/\Lambda$ is small.
For $(0<)s< z$, ${\bf A}({\bf r},z)$ is approximated as
\begin{align}
{\bf A}({\bf r},z) &= -\frac{\phi_0}{4\pi}\int d^2q e^{i{\bf q}\cdot{\bf r}} 
\bigg[ \frac{e^{-qz}+e^{-q(z-s)}}{\Lambda q+1+e^{-qs}}{\bf F}_{\Phi}({\bf q})
\nonumber\\
& ~~~~~~~~~~ -\frac{se^{-q(z-s)}}{s+\Lambda}{\bf F}_{\varphi}({\bf q}) \bigg].
\end{align}
Similarly ${\bf A}({\bf r},z)$ is written as
\begin{align}
{\bf A}({\bf r},z) &= -\frac{\phi_0}{4\pi}\int d^2q e^{i{\bf q}\cdot{\bf r}} 
\bigg[ \frac{e^{-qz}+e^{-q(z-s)}}{\Lambda q+1+e^{-qs}}{\bf F}_{\Phi}({\bf q})
\nonumber\\
& ~~~~~~~~~~ +F_{\Lambda}(z){\bf F}_{\varphi}({\bf q}) \bigg],
\end{align}
where
\begin{eqnarray}
F_{\Lambda}(z)= \left\{
\begin{array}{cc}
-\frac{s}{\Lambda+s}e^{-q(z-s)} &  (s<z) \\
\frac{s-2z}{\Lambda+s} & (0<z<s) . \\
\frac{s}{\Lambda+s}e^{qz} & (z<0)  \\
\end{array}
\right.
\end{eqnarray}
When $0\leq z\leq s$, we have
\begin{equation}
{\bf A}({\bf r},z) \simeq -\frac{\phi_0}{4\pi}\int_0^{\infty} 
\frac{e^{-q|z|}+e^{-q|z-s|}}{\Lambda q+1+e^{-qs}}J_{1}(qr){\bf e}_{\phi}
-\frac{\phi_0}{4\pi}F_{\Lambda}\nabla\varphi.
\end{equation}
When we decompose $\nabla\varphi$ as 
$\nabla\varphi= (\nabla\varphi)_{\varphi}{\bf e}_{\phi}+(\nabla\varphi)_r{\bf e}_r$,
we obtain components $A_{\phi}({\bf r},z)$ and $A_r({\bf r},z)$.
The magnetic field ${\bf B}$ for $0\leq z\leq s$ is simply given as
\begin{align}
B_z({\bf r},z) &= -\frac{\phi_0}{4\pi}\int_0^{\infty}dq
\frac{e^{-q|z|}+e^{-q|z-s|}}{\Lambda q+1+e^{-qs}} qJ_0(qr) \nonumber \\
&~~ -\frac{\phi_0}{4\pi}F_{\Lambda}\frac{1}{r}\bigg[ 
\frac{\partial}{\partial r}(r(\nabla\varphi)_{\phi})-\frac{\partial}{\partial\phi}
(\nabla\varphi)_r \bigg],
\end{align}
\begin{align}
B_r({\bf r},z) &= \frac{\phi_0}{4\pi}\int_0^{\infty}dq
\frac{J_1(qr)}{\Lambda q+1+e^{-qs}}\frac{\partial}{\partial z}
(e^{-q|z|}+e^{-q|z-s|}) \nonumber\\
&~~ +\frac{\phi_0}{4\pi}\frac{\partial F_{\Lambda}}{\partial z}(\nabla\varphi)_{\phi},
\end{align}
\begin{equation}
B_{\phi}({\bf r},z)= \frac{\partial A_r}{\partial z}
= \frac{\phi_0}{4\pi}\frac{2}{\Lambda+s}\frac{\partial\varphi}{\partial r}.
\end{equation}
The second term in $B_z$ vanishes when $\varphi$ is non-singular because of
the identity ${\rm rot}{\rm grad}=0$.

\subsubsection{HFQV in non-equivalent two layers}

Let us examine the magnetic field distribution when two layers are not equivalent
in the presence of HFQV.
We define
\begin{equation}
\frac{2}{\Lambda_1}=\frac{d}{\lambda_1^2},~~~~\frac{2}{\Lambda_2}=\frac{d}{\lambda_2^2}.
\end{equation}
From similar calculations we obtain
\begin{align}
{\bf A}({\bf q},z)&= \bigg[ -e^{-q|z|}(\Lambda_2q+1)+e^{-q|z-s|-qs} \bigg]
\nonumber\\
& \frac{1}{(\Lambda_1q+1)(\Lambda_2q+1)-e^{-2qs}}\frac{\phi_0}{2\pi}
~{\bf F}_{\Phi}({\bf q}),
\end{align}
and we have
\begin{align}
A_{\phi}({\bf r},z)&= \phi_0\frac{1}{2\pi}\int_0^{\infty}dq 
\bigg[ -e^{-q|z|}(\Lambda_2q+1)+e^{-q|z-s|-qs} \bigg]
\nonumber\\
& \frac{1}{(\Lambda_1q+1)(\Lambda_2q+1)-e^{-2qs}}J_1(qr).
\end{align}
When we assume $\Lambda_1\gg s$ and $\Lambda_2\gg s$, in the limit
$r\rightarrow\infty$, $A_{\phi}(r,z=0)$ approaches 
\begin{equation}
A_{\phi}({\bf r},z=0)\simeq -\frac{\Lambda_2}{\Lambda_1+\Lambda_2}
\phi_0\frac{1}{2\pi r}.
\end{equation}
Then the total flux is given by
\begin{equation}
\Phi_{total}(z=0)= \frac{\Lambda_2}{\Lambda_1+\Lambda_2}\phi_0.
\end{equation}
The total flux deviates from half-flux $\phi_0/2$.
When the kink structure of phases in two layers are exchanged, we have
\begin{equation}
A_{\phi}({\bf r},z=0)\simeq -\frac{\Lambda_1}{\Lambda_1+\Lambda_2}
\phi_0\frac{1}{2\pi r},
\end{equation}
and then
\begin{equation}
\Phi_{total}(z=0)= \frac{\Lambda_1}{\Lambda_1+\Lambda_2}\phi_0.
\end{equation}

\subsection{Magnetic field in a knot soliton}

\subsubsection{A knot soliton in a bilayer superconductor}

In this section we investigate a knot state in a two-component superconductor
shown in Fig. 6.  We have a vortex with a unit quantum flux $\phi_0$ and the
phase soliton (kink) of the phase difference $\varphi$.  
The phase difference $\varphi$ is circular symmetric 
in this model and, for example, is given by\cite{yan18d}
\begin{equation}
\varphi = \pi+2\sin^{-1}\left(\tanh \left(\frac{r-r_0}{\delta}\right)\right)
= 4\tan^{-1}\left( e^{(r-r_0)/\delta} \right),
\end{equation}
where the phase soliton exists at $r=r_0$ and $\delta$ denotes the spread of
kink (assuming $\delta \ll r_0$).
In the limit $\delta\rightarrow 0$ this function reduces to a step function,
and we have $\varphi=0$ inside the soliton and $\varphi=2\pi$ outside it.
If we identify the two ends of the vortex, this model becomes a knot soliton.

\begin{figure}[htbp]
\begin{center}
  \includegraphics[height=3.4cm]{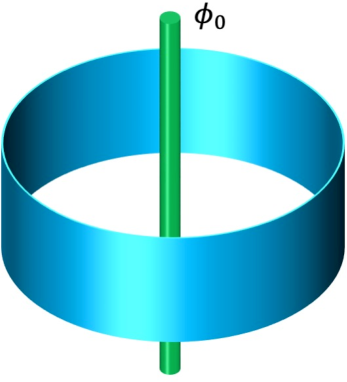}
\caption{Knot-soliton structure in a superconducting bilayer with a unit
quantum flux $\phi_0$ and the phase soliton (kink).
The phase difference $\varphi$ takes the value 0 inside the soliton (kink wall)
and $2\pi$ outside the soliton.
}
\label{fig9}       
\end{center}
\end{figure}

\subsubsection{Magnetic field in a knot soliton}

The magnetic field is given by the formulae in the case
${\bf F}_{\Phi}({\bf q})\neq {\bf F}_{\varphi}({\bf q})$.
We adopt that $s/\Lambda\ll 1$ and then the vector potential is written as
\begin{equation}
{\bf A}({\bf r},z) = -\frac{\phi_0}{4\pi}\int d^2q e^{i{\bf q}\cdot{\bf r}} 
\bigg[ \frac{e^{-qz}+e^{-q(z-s)}}{\Lambda q+1+e^{-qs}}\tilde{{\bf F}}_{\Phi}({\bf q})
+F_{\Lambda}(z){\bf F}_{\varphi}({\bf q}) \bigg],
\end{equation}
where $\tilde{{\bf F}}_{\Phi}({\bf q})=2{\bf F}_{\Phi}({\bf q})$ is for the unit
flux $\phi_0$.
For $0\le z\le s$, the components of ${\bf A}({\bf r},z)$ are given by
\begin{align}
A_{\phi}({\bf r},z) &= -\frac{\phi_0}{2\pi} \int_0^{\infty} dq
\frac{e^{-qz}+e^{q(z-s)}}{\Lambda q+1+e^{-qs}} J_1(qr), \\
A_r({\bf r},z) &= -\frac{\phi_0}{4\pi} F_{\Lambda}(z)
\frac{2}{\delta\cosh((r-r_0)/\delta)},
\end{align}
and $A_z=0$.
Then for the magnetic field ${\bf B}$ we have
\begin{align}
B_z({\bf r},z) &= -\frac{\phi_0}{2\pi}\int_0^{\infty} dq
\frac{ e^{-qz}+e^{q(z-s)} }{\Lambda q+1+e^{-qs}} qJ_0(qr),\\
B_r({\bf r},z) &= -\frac{\phi_0}{2\pi}\int_0^{\infty} dq
\frac{ e^{-qz}-e^{q(z-s)} }{\Lambda q+1+e^{-qs}} qJ_1(qr),\\
B_{\phi}({\bf r},z) &= \frac{\phi_0}{2\pi}\frac{1}{\Lambda+s} 
\frac{2}{\delta\cosh((r-r_0)/\delta)}.
\end{align}
We show ${\bf B}$ as a function of $r/\Lambda$ on the superconducting layer at $z=0$
in Fig. 7.


\begin{figure}[htbp]
\begin{center}
  \includegraphics[height=5.2cm]{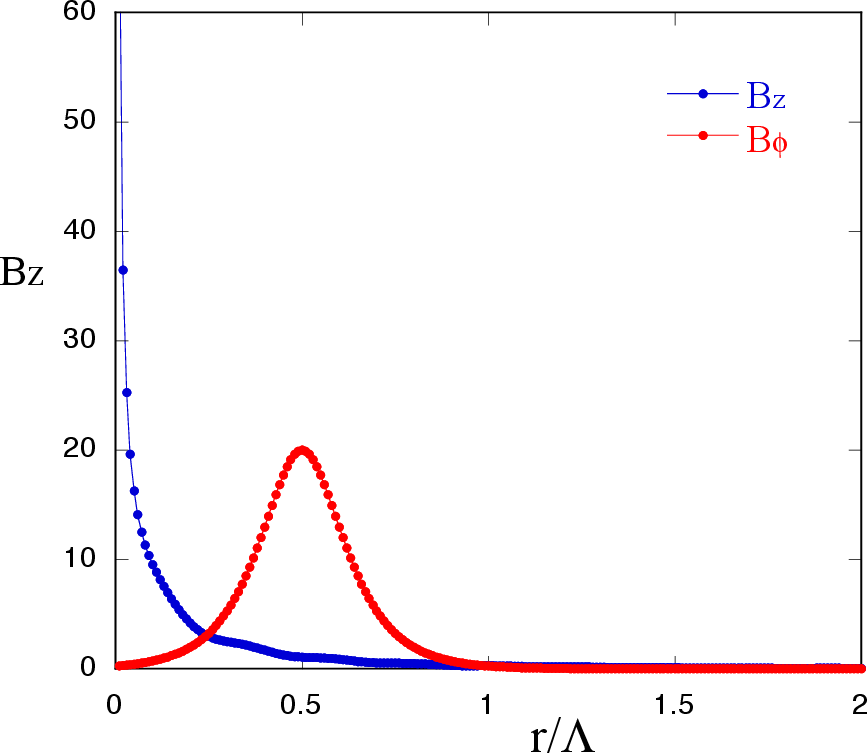}
\caption{Magnetic fields $B_z$ and $B_{\phi}$ in a knot-soliton state
as a function of $r/\Lambda$.
We set $z=0$, $s/\Lambda=0.25$, $r_0/\Lambda=0.5$ and $\delta/\Lambda=0.1$.
}
\label{fig10}       
\end{center}
\end{figure}

\section{Incomplete Meissner effect in a compactified bilayer superconductor}

\subsection{A superconducting bilayer disc}

The currents are written in the form
\begin{align}
{\bf J}_1 &= \frac{e^*}{m_1}\rho_1^2\hbar
\left( \nabla\varphi_1-\frac{e^*}{\hbar c}{\bf A}_1 \right)=\rho_1^2e^*{\bf v}_1,\\
{\bf J}_2 &= \frac{e^*}{m_2}\rho_2^2\hbar 
\left( \nabla\varphi_2-\frac{e^*}{\hbar c}{\bf A}_2 \right)=\rho_2^2e^*{\bf v}_2.
\end{align}
The velocities ${\bf v}_1$ and ${\bf v}_2$ are defined in these equations.
In the polar coordinate, ${\bf J}$ are decomposed into
${\bf J}_{\ell}= J_{\ell r}{\bf e}_r+J_{\ell\phi}{\bf e}_{\phi}$, 
for $\ell=1,2$ (two layers).
In a small disc,  we may put a constraint as
\begin{equation}
 J_{\ell r}=0, ~~~ \ell=1,2
\end{equation}
since the current in the radial direction should vanish.
This results in
\begin{equation}
A_{1r}= \frac{\hbar c}{e^*}\frac{\partial\varphi_1}{\partial r}, ~~~
A_{2r}= \frac{\hbar c}{e^*}\frac{\partial\varphi_2}{\partial r}.
\end{equation}
We define the difference of ${\bf A}_1$ and ${\bf A}_2$ as
\begin{equation}
{\bf A}_d({\bf r})= {\bf A}_1({\bf r},z=0)-{\bf A}_2({\bf r},z=s).
\end{equation}
Then the radial component of ${\bf A}_d$ is given by
\begin{equation}
\left({\bf A}_d\right)_r = \frac{\hbar c}{e^*}\frac{\partial\varphi}{\partial r}.
\end{equation}
From ${\bf J}_1$ and ${\bf J}_2$, the angular component of ${\bf A}_d$ is
\begin{equation}
\left( {\bf A}_d\right)_{\phi}= -\frac{c}{e^*}\left(m_1{\bf v}_1-m_2{\bf v}_2\right)_{\phi}
+\frac{\hbar c}{e^*}\frac{1}{r}\frac{\partial\varphi}{\partial\phi},
\end{equation} 
where 
\begin{equation} 
{\bf p}_d({\bf r})= m_1{\bf v}_1({\bf r},z=0)-m_2{\bf v}_2({\bf r},z=s) 
\end{equation}
indicates the current difference coming from internal phase structure and
$({\bf p}_d)_{\phi}$ is its angular component.
We define $\tilde{B}_z=(\nabla\times {\bf A}_d)_z$, then
\begin{equation}
\tilde{B}_z = \frac{1}{r}\frac{\partial}{\partial\phi}({\bf A}_d)_r
-\frac{1}{r}\frac{\partial}{\partial r}\left( r({\bf A}_d)_{\phi}\right)
= \frac{c}{e^*}\frac{1}{r}\frac{\partial}{\partial r}\left( r{\bf p}_d \right)_{\phi}.
\end{equation}
Hence, when $\partial (r{\bf p}_d)_{\phi}/\partial r\neq 0$, we have $\tilde{B}_z\neq 0$.
Since
\begin{equation}
B_{1z}=B_{2z}+\tilde{B}_z,
\end{equation}
this indicates
\begin{equation}
B_{1z}\neq B_{2z},
\end{equation}
where $B_{jz}=(\nabla\times {\bf A}_j)_z$.
Thus we have at least $B_{1z}\neq 0$ or $B_{2z}\neq 0$.
This shows that the incomplete Meissner effect may occur in a compactified
bilayer superconductor.

\subsection{Magnetic fields}

The vector potential $A_{\pi}$ is given as
\begin{align}
A_{\phi}({\bf r},z) &= -\frac{\phi_0}{4\pi}\int_0^{\infty} dq
\frac{e^{-q|z|}+e^{-q|z-s|}}{\Lambda q+1+e^{-qs}} J_1(qr) \nonumber\\
&~ - \frac{\phi_0}{4\pi}\int d^2q e^{i{\bf q}\cdot{\bf r}}F_{\Lambda}(z)
({\bf F}_{\varphi}({\bf q}))_{\phi}.
\end{align}
The radial component is given by the constraint, for which we use the following
form for $0\leq z\leq s$:
\begin{equation}
A_r({\bf r},z) = -\frac{\phi_0}{4\pi}F_r(z)\frac{\partial\varphi}{\partial r},
\end{equation}
where 
\begin{equation}
F_r(z)= \frac{s-2z}{s}.
\end{equation}
Then for $0\leq z\leq s$ we have
\begin{align}
B_z({\bf r},z) &= -\frac{\phi_0}{4\pi}\int_0^{\infty} dq
\frac{e^{-q|z|}+e^{-q|z-s|}}{\Lambda q+1+e^{-qs}}qJ_0(qr) \nonumber\\
&~ -\frac{\phi_0}{4\pi r}(F_{\Lambda}(z)-F_r(z))
\frac{\partial^2\varphi}{\partial r\partial\varphi},
\end{align}
\begin{align}
B_r({\bf r},z) &= -\frac{\phi_0}{4\pi}\int_0^{\infty}dq
\frac{e^{-qz}-e^{q(z-s)}}{\Lambda q+1+e^{-qs}}qJ_1(qr) \nonumber\\
&~ +\frac{\phi_0}{4\pi}\frac{\partial F_{\Lambda}(z)}{\partial z}
\frac{\partial\varphi}{\partial r}.
\end{align}

We show $B_z$ as a function of $r$ in Fig. 8 where we use $\varphi=(\phi-\pi)f(r)$
with $f(r)=\pi B^*r^2/\phi_0$ and we put $\pi R^2B^*=\phi_0$.
In this model the constant magnetic field $B^*/2$ penetrates in the superconducting
layer at $z=s$.

\begin{figure}[htbp]
\begin{center}
  \includegraphics[height=5.5cm]{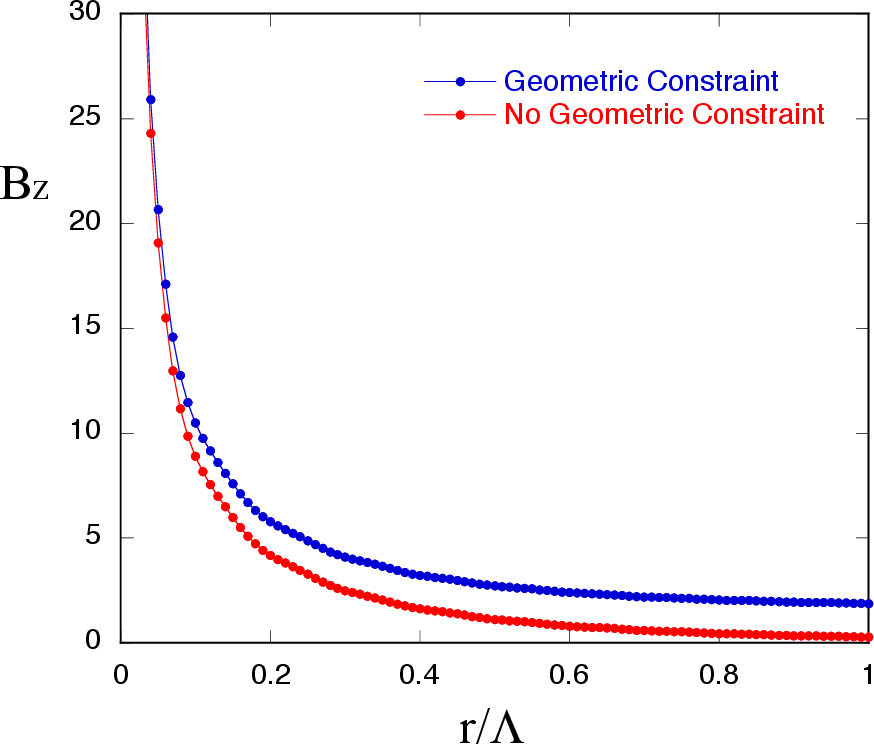}
\caption{Magnetic field $B_z$ in units of $\phi_0/(4\pi\Lambda^2)$ as a function of 
$r/\Lambda$ with geometric 
constraint and without constraint.
We used $s/\Lambda=z/\Lambda=0.25$.  The radius of the disk is
chosen to be $R/\Lambda=1.0$.
}
\label{fig11}       
\end{center}
\end{figure}

\subsection{Current cancellation}

We divide the currents ${\bf J}_{\ell}$ ($\ell=1,2$) into two parts
${\bf J}_{\ell}= {\bf J}_{\ell, {\rm ext}}+{\bf J}_{\ell, {\rm int}}$, 
where ${\bf J}_{\ell, {\rm ext}}$ comes from the applied magnetic field and
${\bf J}_{\ell, {\rm int}}$ is produced by the spatial dependence of phase
structures.  ${\bf J}_{\ell, ext}$ is written as
${\bf J}_{\ell, {\rm ext}}= n_{\ell}e^*{\bf v}_{\ell,{\rm ext}}$, 
for $n_{\ell}=\rho_{\ell}^2$.  ${\bf v}_{\ell,{\rm ext}}$ ($\ell=1,2$)
are velocities induced by the applied external magnetic field for which we have
$m_1{\bf v}_{1,{\rm ext}}= m_2{\bf v}_{2,{\rm ext}}$.
${\bf J}_{\ell,{\rm int}}$ is related to ${\bf v}_{\ell,{\rm int}}$ as
\begin{equation}
{\bf J}_{\ell,{\rm int}}= n_{\ell}e^*{\bf v}_{\ell,{\rm int}}.
\end{equation}
Since internal currents are canceled out each other, we have
\begin{equation}
{\bf J}_{1, {\rm int}}+{\bf J}_{2,{\rm int}}=0.
\end{equation}
Total currents ${\bf J}_{\ell}$ ($\ell=1,2$) are the sum of two contributions:
\begin{equation}
{\bf J}_{\ell}= {\bf J}_{\ell, {\rm ext}}+{\bf J}_{\ell, {\rm int}}
= n_{\ell}e^*{\bf v}_{\ell}.
\end{equation}
The constraint that the current in the radial direction should vanish means
that the external and internal currents cancel each other:
\begin{equation}
({\bf J}_{\ell, {\rm ext}})_r+({\bf J}_{\ell,{\rm int}})_r =0.
\end{equation}
${\bf J}_2={\bf J}_{2,{\rm ext}}-{\bf J}_{1,{\rm int}}= (m_1n_2/m_2n_1){\bf J}_{1,{\rm ext}}-{\bf J}_{1,{\rm int}}$ 
is written as
\begin{equation}
{\bf J}_2
= \frac{m_1n_2}{m_2n_1}{\bf J}_1
-\left(\frac{m_1n_2}{m_2n_1}+1\right){\bf J}_{1,{\rm int}}.
\end{equation}
Since ${\bf J}_{\ell}=n_{\ell}e^*{\bf v}_{\ell}$ ($\ell=1,2$), we obtain
\begin{equation}
(m_1{\bf v}_1-m_2{\bf v}_2)_{\phi} = \frac{m_2n_1}{n_2}
\left(\frac{m_1n_2}{m_2n_1}+1\right)({\bf v}_{1,{\rm int}})_{\phi}.
\end{equation}
Then the momentum difference ${\bf p}_d$ is given by the internal velocity
${\bf v}_{1,{\rm int}}$, that is, the internal current ${\bf J}_{1,{\rm int}}$.
Hence we have the unusual magnetic field $\tilde{B}_z\neq 0$ when
\begin{equation}
\partial (r({\bf v}_{1,{\rm int}})_{\phi})/\partial r\neq 0.
\end{equation}
This indicates that the anomalous field may appear from the non-trivial
angular and radial dependences of the internal current ${\bf J}_{\ell,{\rm int}}$.

\subsection{Free energy}

We argue that the radial component of the field ${\bf A}$ becomes massless
in the presence of anomalous field due to the geometric constraint.
The free energy is written as
\begin{align}
f&= \sum_{\ell}\alpha_{\ell}\rho_{\ell}^2
+\frac{1}{2}\sum_{\ell}\beta_{\ell}^2\rho_{\ell}^4
+\sum_{\ell}\frac{\hbar^2}{4m_{\ell}}(\nabla\rho_{\ell})^2\nonumber\\
&+ \sum_{\ell}\frac{\hbar^2}{4m_{\ell}}
\Biggl\{ \left(\frac{\partial\varphi_{\ell}}{\partial r}\right)^2
+\frac{1}{r^2}\left(\frac{\partial\varphi_{\ell}}{\partial \phi}\right)^2\Biggr\}
\nonumber\\
&- \frac{2e^*}{\hbar c}\sum_{\ell}\frac{\hbar^2}{4m_{\ell}}\rho_{\ell}^2
\left( \frac{\partial\varphi_{\ell}}{\partial r}A_r
+\frac{1}{r}\frac{\partial\varphi_{\ell}}{\partial\phi}A_{\phi}\right)
\nonumber\\
&+ \left(\frac{e^*}{\hbar c}\right)^2\sum_{\ell}\frac{\hbar^2}{4m_{\ell}}\rho_{\ell}^2
(A_r^2+A_{\phi}^2)+\frac{1}{8\pi}(\nabla\times {\bf A})^2 \nonumber\\
&- 2J_{12}\rho_1\rho_2\cos(\varphi_1-\varphi_2).
\end{align}
We impose the gauge condition that the current in the radial direction vanishes:
$\partial\varphi_{\ell}/\partial r-(e^*/\hbar c)A_{\ell r}=0$.
We have the free energy given as
\begin{align}
f&= \sum_{\ell}\alpha_{\ell}\rho_{\ell}^2
+\frac{1}{2}\sum_{\ell}\beta_{\ell}^2\rho_{\ell}^4
+\sum_{\ell}\frac{\hbar^2}{4m_{\ell}}(\nabla\rho_{\ell})^2\nonumber\\
&+ \sum_{\ell}\frac{\hbar^2}{4m_{\ell}}\rho_{\ell}^2\frac{1}{r^2}
\left( \frac{\partial\varphi_{\ell}}{\partial\phi}\right)^2 
- \frac{2e^*}{\hbar c}\sum_{\ell}\frac{\hbar^2}{4m_{\ell}}\rho_{\ell}^2
\frac{1}{r}\frac{\partial\varphi_{\ell}}{\partial\phi}A_{\ell\phi}
\nonumber\\
&+ \left(\frac{e^*}{\hbar c}\right)^2\sum_{\ell}\frac{\hbar^2}{4m_{\ell}}
\rho_{\ell}^2A_{\ell\phi}^2+\frac{1}{8\pi}(\nabla\times {\bf A})^2
\nonumber\\
&- 2J_{12}\rho_1\rho_2\cos(\varphi_1-\varphi_2).
\end{align}
The mass term for $A_{\ell r}$ vanishes and $A_{\ell\phi}$ remains massive.
The massless mode $A_{\ell r}$ results in the incomplete Meissner effect.

\subsection{Discussion on the incomplete Meissner effect in a bilayer
superconductor}

We give a discussion on the incomplete Meissner effect and examine its possibility.
An external magnetic field in a Type-II superconductor is cancelled by the magnetic
field produced by electric currents flowing near its surface.
The magnetic field induced by surface currents is usually uniform.
In a bilayer superconductor the internal field will appear when the phase
difference has spatial dependence.  In this case the internal magnetic field will have
spatial dependence and is inhomogeneous and thus a compensation of the magnetic field by
surface currents may be incomplete.

\section{Summary}

We investigated the magnetic field distribution in a bilayer superconductor
and in a one-layer two-component superconductor.
We have taken account of a phase kink structure as well as the half-flux vortex and
a knot soliton.
We examined the magnetic field originating from the derivative $\nabla\varphi$,
and examined the incomplete Meissner effect and incomplete
Higgs phenomenon in a compactified two-component superconductor with
geometric constraints.
This phenomenon could be called the anomalous Meissner effect.
We investigated a model where the current in one direction is suppressed
extremely due to some geometric constraints.
We considered a disc and have shown that the incomplete Meissner effect may occur
when a non-uniform phase-difference distribution is realized.
This non-uniform phase distribution on a layer could be realized as a kink (soliton)
of the phase difference mode.

The existence of massless modes is also significant in a superconducting bilayer
investigated in the paper.
There is a possibility that a massless mode can be found in this system
with geometric constraints such as a disk.
In contrast, in a one-layer two-component superconducting disk, the constraint
$J_r=0$ is approximately satisfied and we have a small unusual field distribution.

If we take into account a possibility that the phase difference mode will 
produce non-uniform field distributions,
the incomplete Meissner effect will be present in a bilayer superconductor,
where the incomplete Meissner effect indicates that the magnetic field
penetrates into a superconductor partially.  
We discussed this possibility in this paper.
Inhomogeneous magnetic fields are not necessarily compensated by surface currents.
The incomplete Meissner effect is associated with the presence of a massless
component of gauge fields, which is so called the incomplete Higgs mechanism.

Unconventional magnetic distributions are, however, expected to be unstable and 
not to be the ground state.  There have been experimental efforts to observe
half-flux quantum vortices and
unconventional magnetic field distributions in bilayer
superconductors.\cite{tan18c,tan19,tan21,ish22,ish23}  Singular field
distributions could be captured by precise measurements of magnetic field in 
multi-component superconductors.

The numerical calculations were carried out on Yukawa-21 at Yukawa Institute for Theoretical 
Physics in Kyoto University.

\end{document}